\documentclass{jps-cp}

\usepackage{braket}

\newcommand{\pip}{\pi^+}
\newcommand{\pim}{\pi^-}
\newcommand{\pio}{\pi^0}
\newcommand{\pipm}{\pi^\pm}
\newcommand{\alu}{A_{LU}}
\newcommand{\gdiff}{G_1^\perp}
\newcommand{\hdiff}{H_1}
\newcommand{\phih}{\phi_h}
\newcommand{\phir}{\phi_R}
\newcommand{\mgg}{M_{\gamma\gamma}}
\newcommand{\splot}{{\it sPlot}}

\newcommand{\sweights}{{\it sWeights}}
\newcommand{\sweighted}{{\it sWeighted}}
\newcommand{\sweighting}{{\it sWeighting}}

\title{Beam Spin Asymmetries of $\pip\pio$ Dihadrons from SIDIS at CLAS12}

\author{Christopher \textsc{Dilks}$^{1}$ for the CLAS Collaboration}

\inst{
$^{1}$Duke University, Durham, North Carolina 27708, USA
}

\email{christopher.dilks@duke.edu}

\recdate{January 14, 2022}

\abst{
Spin asymmetries provide a wide range of insights into nucleon structure and hadronization.  Recent measurements of beam spin asymmetries of $\pi^+\pi^-$ dihadrons from Semi-Inclusive Deep Inelastic Scattering (SIDIS) at CLAS12 provide the first empirical evidence of a nonzero $G_1^\perp$, the parton helicity-dependent Dihadron Fragmentation Function (DiFF) encoding spin-momentum correlations in hadronization.  These measurements have been extended to aid in the characterization of $H_1^\perp$ and $H_1^\sphericalangle$, the DiFFs dependent on parton transverse spin, via a multidimensional partial wave analysis, giving access to the dependence on the interference of dihadrons of particular polarizations.  Reconstruction of $\pi^0$s allows for further extension of these measurements to $\pi^+\pi^0$ and $\pi^-\pi^0$ dihadrons.  The DiFFs describing $\pi^+\pi^-$ production differ from those describing $\pi^+\pi^0$ and $\pi^-\pi^0$ production, which involve different quark flavors along with a strong suppression of the exclusive diffractive contribution. This presentation will focus on beam spin asymmetries for $\pip\pio$ dihadrons, compared to those from $\pip\pim$ dihadrons, which will help shed light on a more comprehensive picture of dihadron fragmentation.
}

\kword{ dihadron, beam spin asymmetry, SIDIS, CLAS12, dihadron fragmentation function }

\begin{document}
\maketitle

\section{Introduction}
Much understanding of the internal structure and dynamics of the nucleon is gained from high-energy scattering experiments.
Semi-Inclusive Deep Inelastic Scattering (SIDIS) at the CLAS12 detector at Jefferson Lab \cite{Burkert:2020akg} involves the scattering of polarized electrons off a fixed nucleon target. The incoming electron interacts with a quark, effectively knocking it out of the nucleon and causing the formation of outgoing hadrons in a process called hadronization. The kinematic dependence of hadronization is modeled by fragmentation functions; in particular, Dihadron Fragmentation Functions (DiFFs) describe the formation of a pair of hadrons, for example, a $\pip\pim$ dihadron. Spin asymmetries in dihadron production, such as the presented measurement, together with known Parton Distribution Functions (PDFs), provide constraints on the kinematic dependences of DiFFs, offering insight into the hadronization process in general.

The CLAS12 detector has been used to measure the beam spin asymmetry in the $ep\to e\pip\pim X$ SIDIS process \cite{Hayward:2021psm}. The beam spin asymmetry $\alu$ is written
\begin{equation}
\alu=\frac{1}{P_B}\frac{N^+\left(\phih,\phir\right)-N^-\left(\phih,\phir\right)}{N^+\left(\phih,\phir\right)+N^-\left(\phih,\phir\right)},
\end{equation}
where $N^\pm$ is the dihadron yield from scattering a $\pm$ helicity electron, with helicity defined in the proton rest frame, and $P_B$ is the electron beam polarization. $N^\pm$ is a function of $\phih$ and $\phir$, the dihadron azimuthal angles \cite{Hayward:2021psm,Bacchetta:2002ux}. Sinusoidal functions of $\phih$ and $\phir$ modulate $\alu$, and each corresponding modulation amplitude is sensitive to a particular PDF coupled with a DiFF. At leading twist, $\alu$ is sensitive to the unpolarized PDF $f_1(x)$ coupled with $\gdiff$, the DiFF dependent on the longitudinal spin of the fragmenting quark \cite{Bacchetta:2002ux,Matevosyan:2017liq}. With $f_1(x)$ well-constrained \cite{Bacchetta:2019sam}, twist-2 $A_{LU}$ modulations constrain $\gdiff$. At twist 3 the sensitivity is to the twist-3 PDF $e(x)$ coupled with $H_1^\perp$ and $H_1^\sphericalangle$, collectively denoted by $\hdiff$, the DiFFs dependent on the transverse spin of the fragmenting quark \cite{Bacchetta:2003vn}. With $\hdiff$ constraints from Belle \cite{Belle:2011cur,Courtoy:2012ry}, the recent $\alu$ measurement \cite{Hayward:2021psm} can help provide a point-by-point extraction of $e(x)$ \cite{Courtoy:2014ixa}.

The presented measurement complements the $\pip\pim$ measurement \cite{Hayward:2021psm} with $\pip\pio$ dihadrons. The $\pip\pim$ channel is sensitive to DiFFs for the fragmentation of quarks to $\pip\pim$, however isospin and charge relations imply
$
D^{u\to\pip\pim}=
D^{\bar{d}\to\pip\pim}=
D^{d\to\pim\pip}=
D^{\bar{u}\to\pim\pip},
$
where $D$ is any DiFF \cite{Bacchetta:2006un}. On the other hand, $\pip\pio$ and $\pim\pio$ dihadrons access DiFFs with different flavor dependence, independent of DiFFs for $\pip\pim$. Given constraints on $f_1(x)$ and $e(x)$, the presented measurement of $\pip\pio$ dihadrons offers a glimpse into the behavior of $\gdiff$ and $\hdiff$ for fragmentation into $\pipm\pio$ dihadrons.

\section{Analysis}
The observation of $\pip\pio$ dihadrons requires the reconstruction of $\pio$s from photon pairs. Thus semi-inclusive events with the final state $e^-\pip\gamma\gamma X$ are chosen, with the detected particles in the CLAS12 Forward Detector \cite{Burkert:2020akg}, between scattering angles of $5^\circ$ and $35^\circ$. The electron beam energy ranged from 10.2--10.6~GeV, with a polarization of 86--89\%, and was scattered on a liquid hydrogen target. 

Event selection criteria are similar to those from Ref.~\cite{Hayward:2021psm}. Inclusive cuts are $Q^2>1~\text{GeV}^2$, $W>2$~GeV, and $y<0.8$. Both pions must have positive Feynman-$x$ and the dihadron must have $z<0.95$. For accurate track reconstruction, the $\pip$ momentum must be above $1.25$~GeV, and the minimum allowed photon energy is $0.6$~GeV to be above calorimeter noise levels. The photons are also required to have $\beta$, the ratio of measured particle velocity to the speed of light, to be within $0.9<\beta<1.1$, and to be at least $8^\circ$ from the scattered electron. For this dihadron channel, and unlike $\pip\pim$, there is no evidence of exclusive dominance at low missing mass $M_X$, thus no cut is applied on $M_X$. Finally, additional refinement cuts are used for the particle identification, vertex, and fiducial volume.

\begin{figure}[t]
\centering
\includegraphics[width=0.4\textwidth]{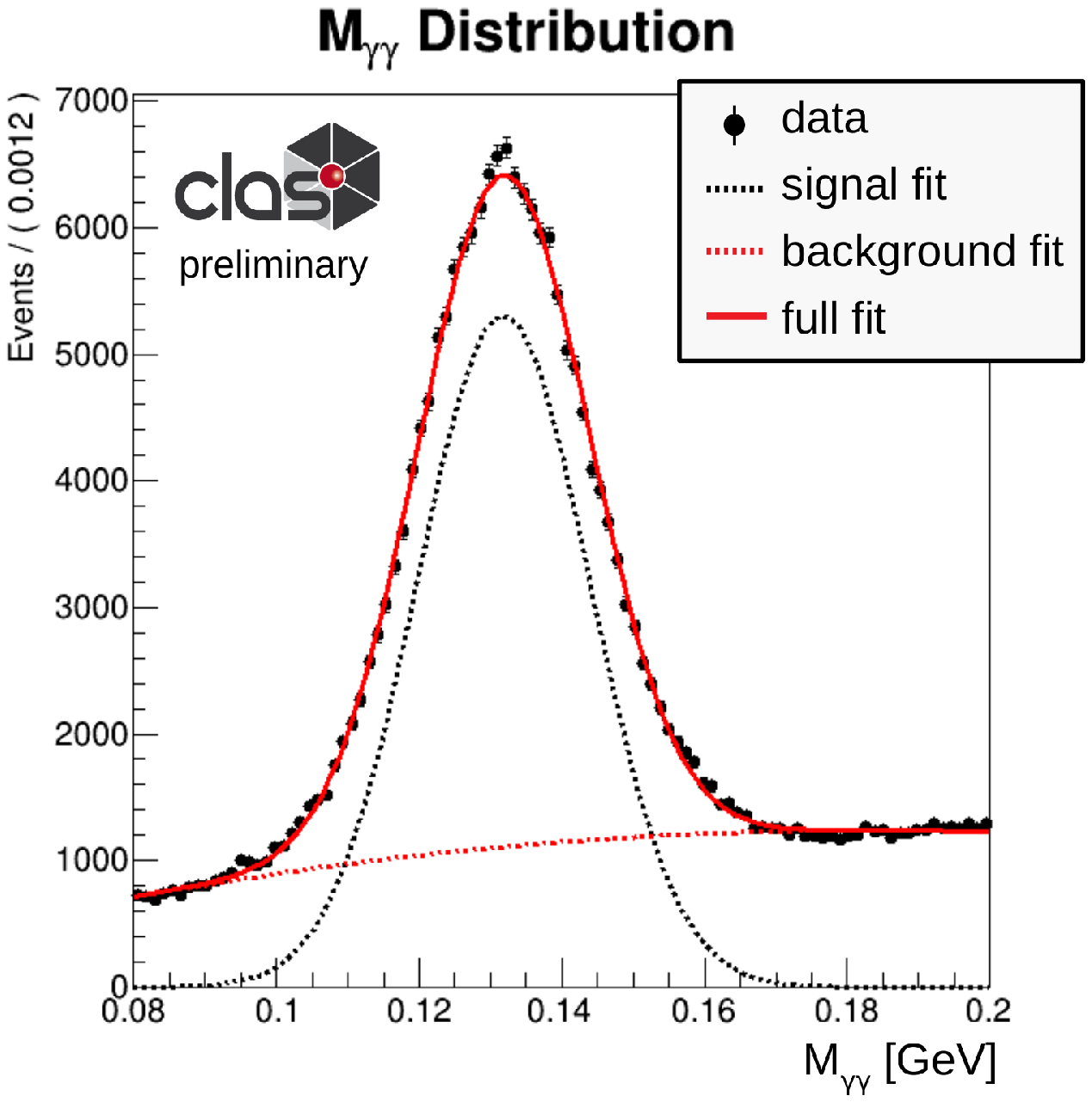}
\caption{Diphoton mass for a selected $\alu$ measurement bin. The black points are data, and the solid red curve is the fit, composed of the signal (black, dashed) and background (red, dashed) fits.}
\label{fig:diphM}
~\\
\includegraphics[width=0.45\textwidth]{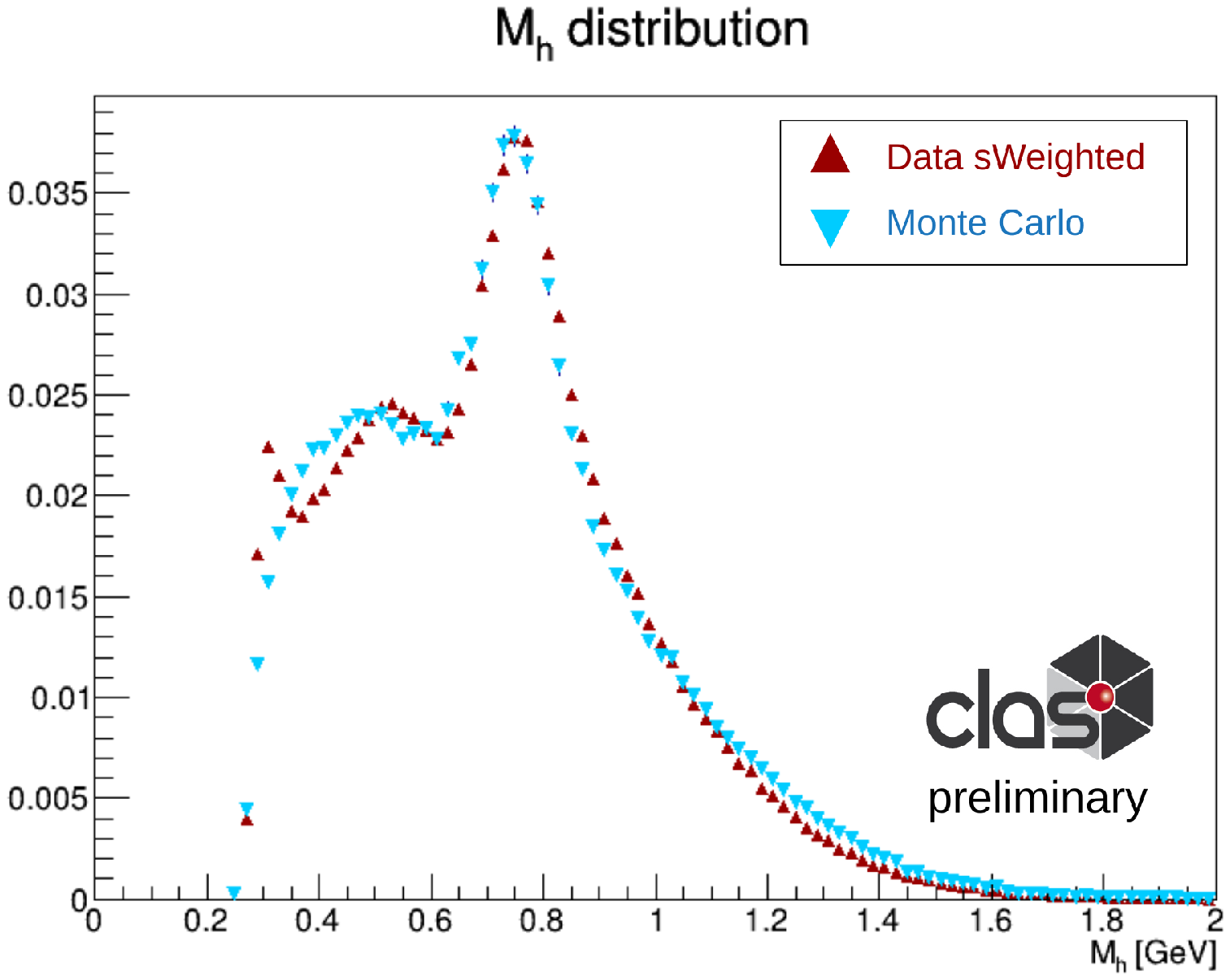}
\includegraphics[width=0.45\textwidth]{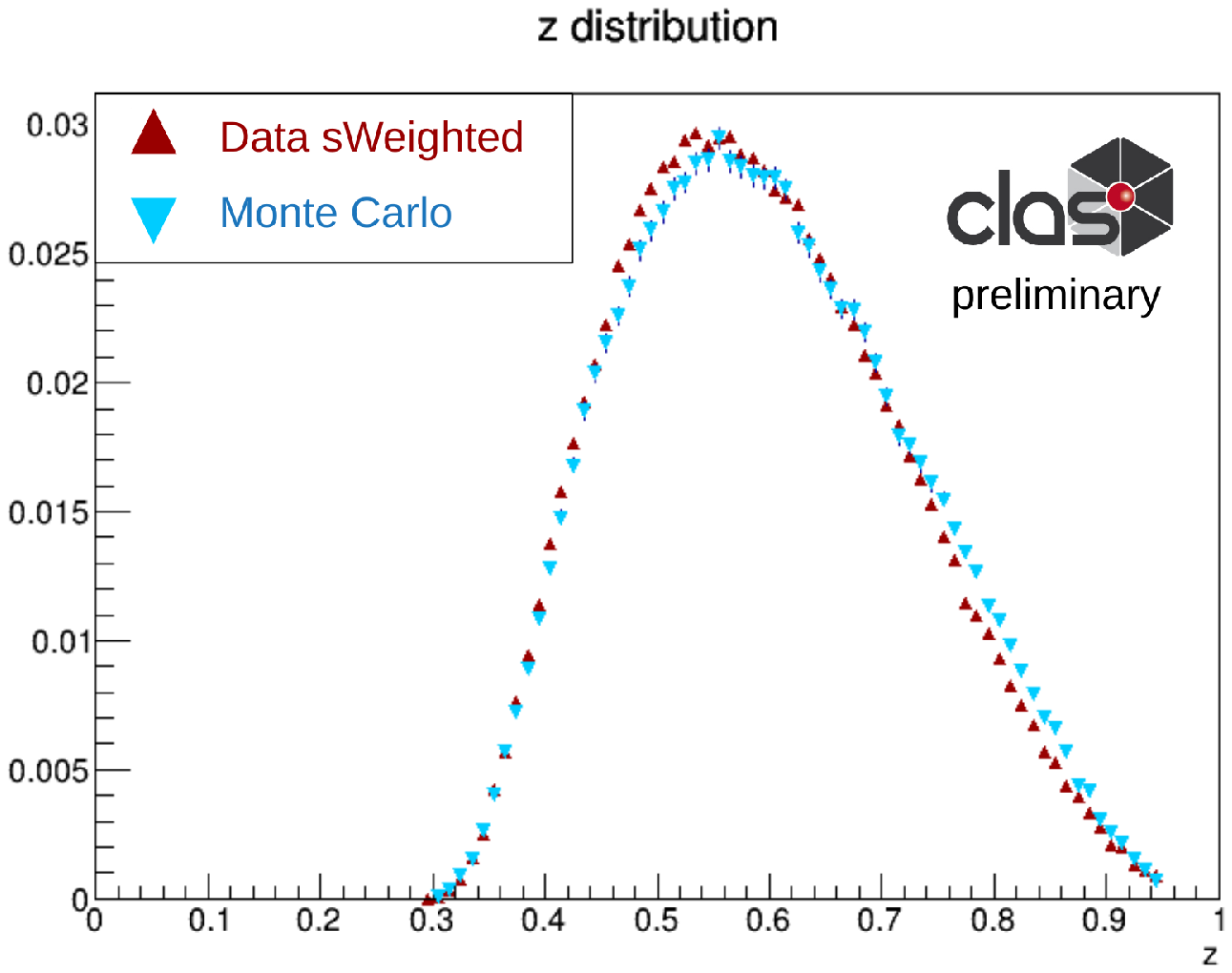}
\caption{Distributions of $\pip\pio$ $M_h$ (left) and $z$ (right). The red upward triangles denote the \sweighted~data distribution, and the blue downward triangles denote the MC distribution, reconstructed from true $\pip\pio$ dihadrons. The MC data are not \sweighted, and all distributions are normalized by the sum of their weights.}
\label{fig:sweighted}
\end{figure}

The $\pio$s are reconstructed from semi-inclusive photon pairs, diphotons. Figure \ref{fig:diphM} shows the diphoton invariant mass $\mgg$, where the black points are data, and the curves represent a fit. The $\pio$ mass peak is fit with a Gaussian function, shown as the black dashed curve, on top of background modeled by a quadratic Chebyshev polynomial, shown as the red dashed curve. The full fit result is shown as the solid red curve. Because of the significant background under the $\pio$ peak, a correction of $\alu$ for background contamination is needed.

A diphoton mass fit and $\alu$ background correction is performed for each $\alu$ measurement bin. The \splot~technique \cite{Pivk:2004ty} is used for background correction, which formulates the calculation of weights that provide a statistical separation of signal and background, called \sweights. The \sweights~are calculated for each event using the $\mgg$ fit functions and covariance matrices. Weighting the data distribution of some variable $d$ with $\pio$-signal \sweights~reproduces, on average, the $d$ distribution of the true $\pio$ signal. The efficacy of this technique requires $d$ to not be correlated with $\mgg$, however.

Figure \ref{fig:sweighted} shows example kinematic distributions, the dihadron invariant mass $M_h$ and dihadron $z$, to demonstrate the effectiveness of \sweighting. The red upward triangles denote the \sweighted~$\pip\pio$ distributions from data, while the blue downward triangles denote the reconstructed $\pip\pio$ Monte Carlo (MC) data. These MC data are not \sweighted, since the majority of diphoton background events are excluded by matching to true $\pip\pio$ dihadrons in the generated MC data sample. To allow comparison of the distribution shapes, the \sweighted~data are normalized by the sum of the \sweights, and the MC data are normalized by the number of dihadrons. In general the data and MC distributions agree, and a $\rho^+$ peak is visible around $M_h=0.77$~GeV. There are some small disagreements at low $M_h$, where the diphoton background $\mgg$ correlates with $M_h$ near the edge of the phase space, impacting the \sweighting~accuracy in this region.

An \sweighted-likelihood fit \cite{Langenbruch:2019nwe} is used to extract the background-corrected $\alu$. The same seven $\alu$ modulations as in Ref.~\cite{Hayward:2021psm} are simultaneously fit for their amplitudes, where the modulations are of the two azimuthal angles of dihadrons, $\phih$ and $\phir$. The presented measurement focused on three particular amplitudes:
$\alu^{\sin(\phih-\phir)}$,
sensitive to $f_1G_1^{\ket{\ell,1}}$;
$\alu^{\sin(2\phih-2\phir)}$,
sensitive to $f_1G_1^{\ket{\ell,2}}$;
and
$\alu^{\sin\phir}$,
sensitive to $eH_1^{\perp\ket{\ell,1}}$.
The DiFF superscripts denote terms in the partial wave expansion \cite{Gliske:2014wba}, whereas $\gdiff$ and $\hdiff$ denote the full DiFF sets. Although partial waves are beyond the scope of this presentation, note that $G_1^{\ket{\ell,2}}$ dominantly involves the interference of $p$-wave dihadrons, which includes a large fraction of vector meson decays, in particular $\rho\to\pi\pi$, whereas $G_1^{\ket{\ell,1}}$ involves both $sp$ and $pp$ interference. 

\section{Results}
\begin{figure}[t]
\centering
\includegraphics[width=0.85\textwidth]{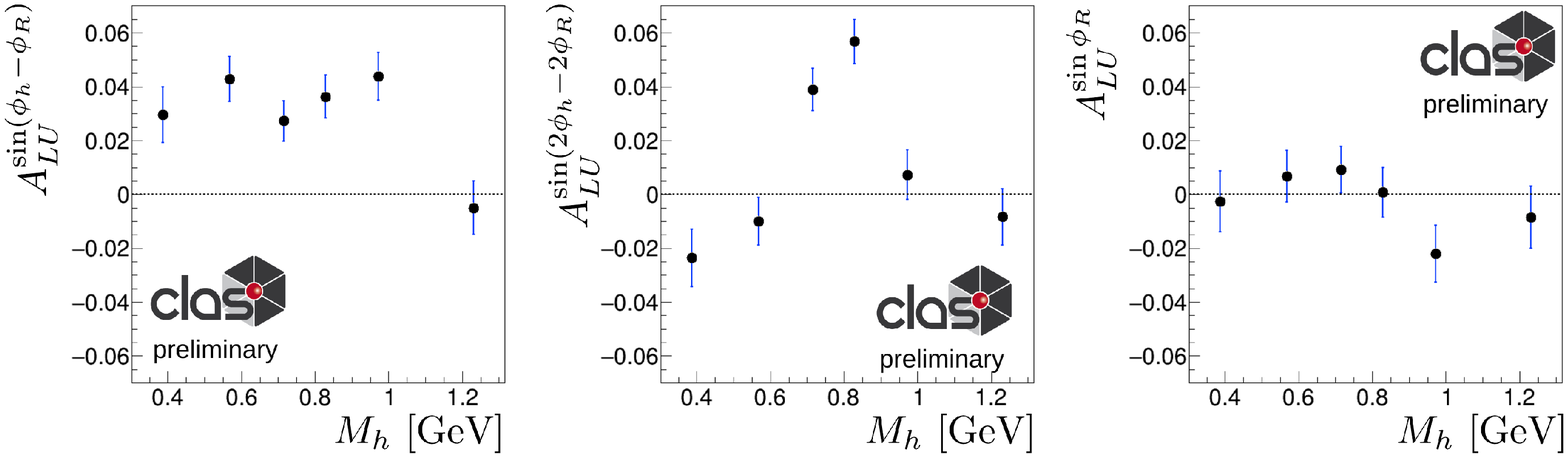}\vspace{0.2cm}
\includegraphics[width=0.85\textwidth]{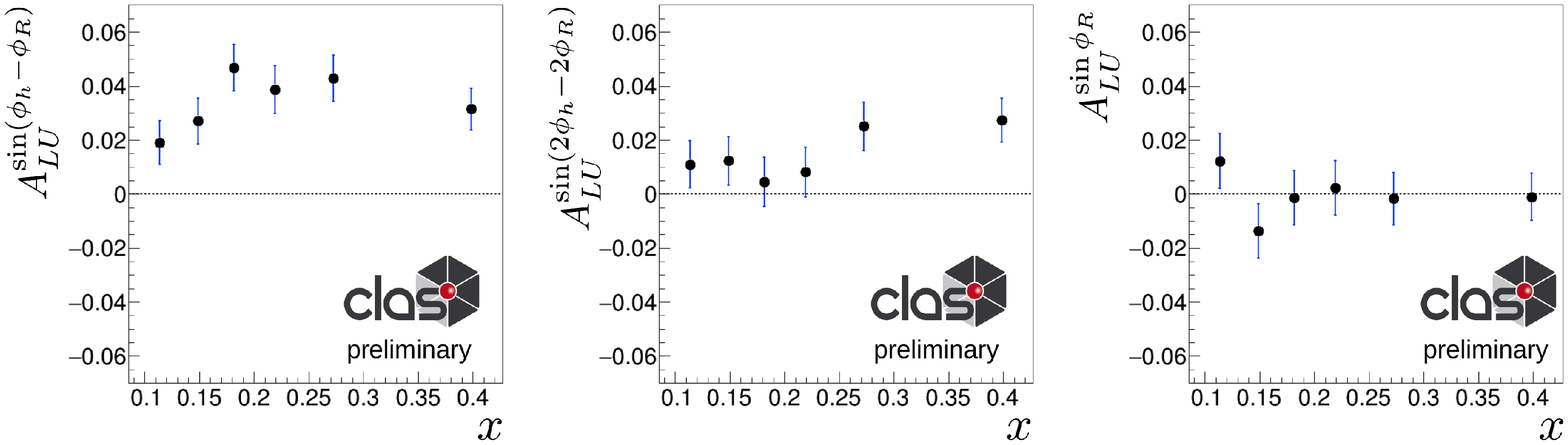}\vspace{0.2cm}
\includegraphics[width=0.85\textwidth]{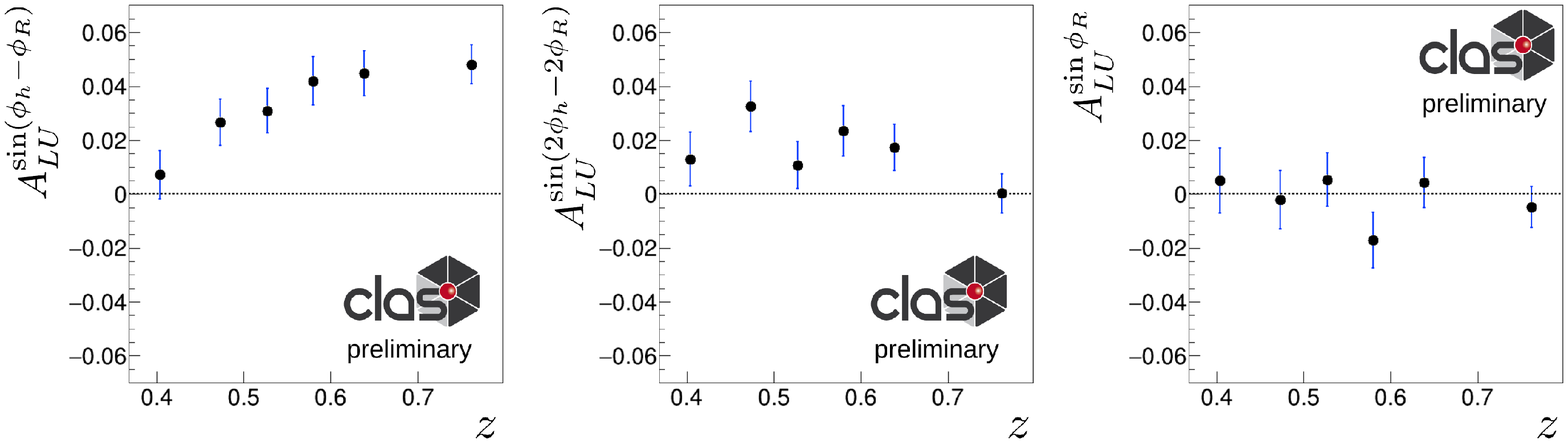}
\caption{Preliminary measurements of beam spin asymmetries in SIDIS $\pip\pio$ production at CLAS12, in bins of $M_h$ (top row), $x$ (middle), and $z$ (bottom). Three amplitudes are shown: 
$\alu^{\sin(\phih-\phir)}$ (left column),
$\alu^{\sin(2\phih-2\phir)}$ (middle), and
$\alu^{\sin\phir}$ (right).
}
\label{fig:asym}
\end{figure}

Figure \ref{fig:asym} presents preliminary measurements of $\alu$ for $\pip\pio$ dihadrons. The three rows show the $M_h$, $x$, and $z$ dependences, while the three columns show the three aforementioned amplitudes. The first amplitude, $\alu^{\sin(\phih-\phir)}$, sensitive to $f_1G_1^{\ket{\ell,1}}$, is positive for most the kinematic range. The $\pip\pim$ dihadron $\alu^{\sin(\phih-\phir)}$ sign-change feature at $M_h{\sim}M_\rho\approx 0.77$~GeV \cite{Hayward:2021psm} does not appear in these $\pip\pio$ measurements.
The asymmetry increases as a function of $z$, as expected (see Ref.~\cite{Matevosyan:2017alv}), but it is difficult to discern if the $\pip\pio$ $\alu$ is smaller than the $\pip\pim$ $\alu$ magnitude, because of the aforementioned sign change.
The difference of mass dependences between the two dihadron channels indicates a decay-product flavor dependence of $\gdiff$.

The second amplitude, $\alu^{\sin(2\phih-2\phir)}$, sensitive to $f_1G_1^{\ket{\ell,2}}$, shows a distinct enhancement around $M_\rho$. This enhancement was also observed in the $\pip\pim$ measurement \cite{Dilks:2021nry}, and seems to be a typical feature of the $pp$-interference DiFF partial wave \cite{Luo:2020axe,Courtoy:2012ry}.

The third amplitude, $\alu^{\sin\phir}$, sensitive to $eH_1^{\perp\ket{\ell,1}}$, is consistent with zero, unlike the significantly positive amplitude seen in $\pip\pim$. Assuming $e(x)$ is the same for both processes, the difference between these measurements is caused by a difference in $H_1$ between the dihadron channels. 

\section{Conclusions}
Measurements of $\alu$ from $\pip\pio$ dihadrons extend previous measurements from $\pip\pim$ dihadrons \cite{Hayward:2021psm} and grant access to different sets of channel-dependent DiFFs. While the $\pip\pio$ channel appears to have a suppression of exclusive events, when compared to the $\pip\pim$ channel, the involvement of $\pio\to\gamma\gamma$ decays requires a background correction implementation. The \sweighting~technique is used to perform a background-corrected fit for the extraction of $\alu$ amplitudes. Significant twist-2 amplitudes, sensitive to $\gdiff$, are observed, with no sign change at the $\rho^+$ mass. The twist-3 amplitude, sensitive to $e(x)$ and $\hdiff$, is consistent with zero. We plan to extend this analysis to the $\pim\pio$ channel, to provide a more complete picture. Furthermore, assessment of systematic uncertainties is needed, which will differ from Ref.~\cite{Hayward:2021psm}, given the involvement of $\pio$ reconstruction. It also may be possible to fit the partial waves, which would give further insight into angular momentum correlations \cite{Dilks:2021nry}.

\vspace{0.3cm}
\paragraph{Acknowledgements}
We acknowledge the outstanding efforts of the staff of the Accelerator and the Physics Divisions at Jefferson Lab in making this experiment possible.

\paragraph{Funding information}
This work was supported in part by the U.S. Department of Energy, the National Science Foundation (NSF), the Italian Istituto Nazionale di Fisica Nucleare (INFN), the French Centre National de la Recherche Scientifique (CNRS), the French Commissariat pour l$^{\prime}$Energie Atomique, the UK Science and Technology Facilities Council, the National Research Foundation (NRF) of Korea, the Helmholtz-Forschungsakademie Hessen für FAIR (HFHF) and the Ministry of Science and Higher Education of the Russian Federation. The Southeastern Universities Research Association (SURA) operates the Thomas Jefferson National Accelerator Facility for the U.S. Department of Energy under Contract No. DE-AC05-06OR23177.
The work of CD is supported by the U.S. Department of Energy, Office of Science, Office of Nuclear Physics under Award Number DE-SC0019230.

\end{document}